\documentclass[intlimits,twoside,a4paper]{article}

\usepackage[utf8]{inputenc}
\usepackage{cmpj3}

\usepackage{alphabeta}
\usepackage{dsfont}
\usepackage{graphicx}
\usepackage{bm}

\usepackage{color}

\usepackage{latexsym}
\usepackage{mathrsfs}

\usepackage{amsfonts}
\usepackage{epsfig}
\usepackage{hyperref}

\usepackage{soul}
\usepackage{cancel}
\usepackage[normalem]{ulem}
\usepackage[cmtip,all]{xy} 
\usepackage{comment}
\usepackage{multirow}
\usepackage{mathtools}

\usepackage{mathrsfs}

\usepackage{hhline}

\newcommand{\be}{\begin{equation}}
\newcommand{\ee}{\end{equation}}
\newcommand{\bea}{\begin{eqnarray}}
\newcommand{\eea}{\end{eqnarray}}
\newcommand{\bem}{\begin{matrix}}
\newcommand{\eem}{\end{matrix}}
\newcommand{\nnb}{\nonumber}
\def\nnb{\nonumber}

\issue{2024}{27}{3}{33201}
\doinumber{10.5488/CMP.27.33201}

\title{Differential geometry, a possible avenue for thermal ablation in oncology?}
\author[A. Manapany, 	L. Didier,	L. Moueddene, 	B. Berche,	S. Fumeron]{ A. Manapany\orcid{0000-0002-7138-0659}\refaddr{label1,label2},
	L. Didier\refaddr{label1},
	L. Moueddene\refaddr{label1,label2},
 	B. Berche\orcid{0000-0002-4254-807X}\refaddr{label1,label2}\footnote{Email:\email{bertrand.berche@univ-lorraine.fr}},
	S. Fumeron\refaddr{label1} }
\addresses{
\addr{label1} Laboratoire de Physique et Chimie Th\'eoriques,  CNRS - Universit\'e de Lorraine, UMR 7019,
    	 Nancy, France 
\addr{label2} $\mathbb{L}^4$ collaboration, Leipzig-Lorraine-Lviv-Coventry, Europe
	}
	
\date{Received December 25, 2023}
\begin{document}
\maketitle

\begin{abstract}
We report a model for hyperthermia therapies based on heat diffusion in a biological tissue containing a topological defect. 
Biological tissues  
behave like active liquid crystals with 
the presence of topological defects which are likely to anchor tumors during the metastatic phase of cancer evolution and the therapy challenge is to destroy the cancer cells without damaging  surrounding healthy tissues. 
The defect creates an effective non-Euclidean geometry for low-energy excitations, modifying the bio-heat equation. Applications to protocols of 
thermal ablation for various biological tissues (liver, prostate, and skin) is analyzed and discussed.
   \end{abstract}





\section{The context}
\label{sec:intro}

In essence, cancer consists of an uncontrolled growth of abnormal cells that invade healthy neighboring body parts. Despite important breakthroughs in the understanding, diagnosis, and treatment of the various forms of cancer, it is still the second leading cause of mortality worldwide \cite{WHO2020}. This incentivizes the development of new tools to supplement chemotherapies and radiotherapies, such as {i}mage-guided thermal tumor ablation \cite{ChuDupuy, Brace}. 

Hyperthermia treatments must obey two constraints: on the one hand, they have to destroy the cancer cells without damaging the surrounding healthy tissue \cite{Brace}, and on the other hand, thermal ablation must be performed completely, otherwise the heating of the tumor cells may promote increased tolerance concerning elevated temperatures and even trigger accelerated cancer cells growth \cite{markezana2020moderate}. Hence, the efficiency of thermal ablation protocols requires 1) a thorough understanding of heat diffusion processes in the vicinity of cancer cells and 2) a precise account of the deposited energy, a delicate tuning between the temperature reached and the duration of the treatment.

A possible track to fulfil requirement 1) is to track down topological defects. For the physicist, the notion of topological defect is ubiquitous (phase transitions, cosmology, high energy physics, solid state, liquid crystals, etc. \cite{reviewFB}), but that it could be of probable interest in oncology is like a surprise at first glance. The presence of topological defects in biological tissues is now well-established and their possible role in biological processes soon attracted the attention of physicists \cite{BouligandLesHouches}, but it is only recently that studies have shown how this role may also be central in certain kinds of cancer. Indeed, body organs are covered in the surface by the epithelium, which behaves as an active liquid crystal: the epithelial cells display an orientational order within domains, and at the boundaries between competing domains, topological defects arise \cite{DoostmohammadiAdamerEtAl,doostmohammadi2021physics,Ardaseva}. During the metastatic phase of cancer evolution, cancer cells concentrate in the vicinity of topological defects due to lower cell density in the core region \cite{SawEtAl,zhang2021topological}, which makes them attractive targets for thermal therapies.

In addition to the vast body of available data, hyperthermia may highly benefit from a predictive mathematical model to optimize the treatment strategies. Although cancers are convoluted and multifactorial diseases, such highly complex behaviors may often be captured by a set of elementary physical mechanisms: this is the framework of mathematical oncology, which has become --- although not without some difficulty --- a compulsory tool to enable cancer discoveries \cite{Gatenby2003, Jackson2014, Rockne2019, Bull2022}. In this work, we report a differential geometry-based model designed to characterize and discriminate thermotherapy protocols in oncology. This may have some 
 relevance for image-guided interventional therapies.
We solve numerically Pennes' bioheat equation in a biological tissue displaying a topological defect. The heat balance encompasses the influence of the metabolism and blood flow \cite{Pennes48,AndreozziEtAl,TucciEtAl}, along with the energy input due to the thermal treatment. The defect is included by writing the diffusion equation on a curved manifold \cite{Smerlak1}, since  defects induce  non-Euclidean geometries for propagating low-energy excitations \cite{Katanaev,SatiroMoraes}. Depending on the  physical properties of the 
biological tissue where a tumor is supposed to be present (for instance liver, skin, prostate...), we can thus estimate, in the form of a compromise between the temperature to be applied and the duration of application, the scenario that seems to be the most appropriate.

In this preliminary study, we aim to show that the Physics of topological defects and heat transport in the presence of such defects may have something meaningful to say in the very practical field of cancer therapies but the model proposed here has strong limitations: i)  it is a continuum field theory approach (here in cylindrical symmetry), a coarse-grained description at scales much larger than the cell sizes. ii) The physical parameters describing the tissues are taken from the literature and are supposed to be fixed, i.e., their temperature dependence is neglected. This does not account for the possible response of the organism to the elevation of temperature. iii) The mechanisms of cell death are complex biological processes, which are here over-simplified, considering that after a period determined by the imposed temperature, the cells are killed and that this does not depend on the type of tissue considered. All these aspects should be analyzed in deeper detail to ensure a better validity of the model. These aspects are discussed in the conclusion. 

The paper is structured as follows. First, we describe at a very introductory level recent observations made in medicine and biology --- to be specific we will refer later to the case of ovarian, liver, prostate, and skin cancers, but the observations are of a wider range of applications. We then propose a physical description of a model system that we believe is reasonably close to the experimental situation considered and present the tenets of heat diffusion on non-Euclidean geometries. We finally performed several case studies to determine which hyperthermia protocol seems to be more efficient for maintaining the core of the defect at high temperatures while preserving the surrounding tissues.



\section{Topological defects in biological tissues drive tumors location during the metastatic cascade} 

Here we summarize the present knowledge on the role of topological defects in biological tissues on the mechanisms of pinning and growth of metastases since this will be the building block of our later analysis. 
{A crucial phase in the development of many cancers is indeed the metastasis formation: cancer cells spread from the primary tumor and colonize distant secondary organs.} To do so, cancer cells must successfully go through a series of steps (the metastatic cascade \cite{Pantel2004}):
the cells must detach from a primary tumor ({\sl extrusion}) and invade the surrounding tissue (this is called {\sl EMT}, see the glossary at the end of the paper) \cite{GraciaTheisEtAl}, then enter a blood or lymphatic vessel ({\sl intravasion}) where they circulate to reach distant organs. There, the cancer cells exit the vessel ({\sl extravasion}), and grow in a new tissue. 
This is the metastatic growth. During all these steps, the tumor cells  encounter immune cells to which they must be resistant.

The {\sl Yes-Associated Protein} (YAP)  and the {\sl Transcriptional Co-activator with PDZ-binding Motif} (TAZ) are regulators of early embryonic development and the growth of  tissues. Though they are not direct drivers of cancer, they have been implicated in cancer development and progression due to their involvement in various cellular processes.
They drive the transcription of genes that promote cell proliferation, cell survival, and stem cell maintenance \cite{Warren2018,cheng2022biology}. In vitro and in vivo evidence (in particular with human  cell lines) suggest that inappropriate YAP or TAZ activation can promote metastasis by influencing virtually every process of the aforementioned cascade.
In particular, the authors of reference~\cite{Warren2018} write: 
\begin{quotation}``a meta-analysis of 21 different studies with a combined 2983 patients revealed that YAP is overexpressed and associated with poor outcomes and reduced survival in many human cancers. This is complemented by numerous in vivo studies that show a role for YAP or TAZ in tumor formation and growth'' (\dots) ``Several other [studies] have directly implicated YAP, TAZ, or TEADs in metastasis of numerous cancer types, including melanoma, lung cancer, breast cancer, cholangiocarcinoma, gastric cancer, ovarian cancer, colorectal cancer, and oral squamous cell carcinoma.'' 
\end{quotation}
 Those studies show that overexpression of YAP or TAZ  enhances the morphological changes associated with EMT and promotes cell migration and invasion of {\sl mesothelioma}.  A few studies have also implicated YAP in the intravasion in blood vessels and suggested that YAP and TAZ enhance tumor cell survival in circulation. Cancer cells with high YAP/TAZ activity show resistance to {\sl anoikis} (cell's death) since YAP/TAZ activity in tumor cells also helps these cells to evade  the immune system, which is a critical step for proliferation, tumor progression, and metastasis. 
Authors of  reference~\cite{Warren2018} conclude that 
\begin{quotation}``experimental evidence (\dots)  suggests that YAP/TAZ activation, which occurs in many human cancer types, is pro-tumorigenic and pro-metastatic''. \end{quotation}

Epithelial and mesothelial tissues are typically organized in a highly ordered and regular manner, with cells tightly packed and adhering to one another to form continuous sheets. However, under certain conditions or during specific developmental processes, topological defects or deviations from perfect regularity can occur in these tissues.
Many aggressive cancer cells already have elevated YAP/TAZ activity before they enter circulation, but being exposed to shear stress or disturbed flow induced by the presence of topological defects further activates YAP and TAZ. If the mechanisms leading to defect formation in biological tissues are not fully understood,  correlations have been identified between collagen density and tumorigenesis. Tumor-associated collagen signatures (TACS) are used for the identification of pre-palpable
tumors in tissues rich in collagen. Authors of  reference~\cite{provenzano2006collagen} showed that for TACS-2 tumors (non-invasive), collagen fibers
are stretched parallelwise to the tumor regular boundary, forming a nest around it, whereas for TACS-3 tumors (invasive), collagen fibers align normally to tumor irregular boundary: one may therefore expect that such an orientation transition is accompanied by the germination and growth of defects which can then promote local invasion or migration of epithelial cells. 
In different epithelium types, singularities in the cell alignments were shown to be correlated with  extrusion sites \cite{SawEtAl}, demonstrating that defect-induced anisotropic stresses are the primary precursors of mechanotransductive responses in cells, including YAP transcription factor activity \cite{elosegi,SawEtAl,zhang2021topological,DoostmohammadiAdamerEtAl,doostmohammadi2021physics}. 
In particular,
comet-like topological defects (defects of order $+1/2$, as we will see in the next section) are present on the epithelium where they are essential to cell extrusion processes \cite{GraciaTheisEtAl,HoganEtAl}. 
As the comet generates strong compressive stress at its core, it causes the YAP protein to migrate from the nucleus of the cell to its cytoplasm, causing apoptosis and  extrusion \cite{SawEtAl}.
In all the cases observed, the head of the comet pointed towards the cell to be eliminated. It should be noted that if the defects are responsible for the extrusion, the converse was not observed and their appearance occurs well before the elimination of the cells. {The statistical analysis of cell velocities in the vicinity of comet-like defects also showed a sink effect along the defect axis, with cells on both sides of the defect moving inward (\cite{zhang2021topological}).} 

The trefoil-like defects ($m=-1/2$) constitute a large part of the foci for the development of ovarian cancer metastases on the mesothelium --- a specialized type of epithelium --- of the abdomen. The latter is composed of mesothelial cells and presents the same nematic ordering as the epithelial cells of the ovaries. The study of the mesothelium is of great interest for ovarian cancer since this is where most of the metastases are formed. Unlike comets, trefoil-like defects display a net outward cell velocity (outside their legs), leading to a lower cell density in the core region. {Correlations have been observed between the presence of defects of order $-1/2$ and metastases development, hence reducing the clearance of cancer cells \cite{zhang2021topological}}.

\section{Thermal tumor ablation, a promising treatment}
Delayed diagnosis and treatment of cancer is a major cause of mortality.  Thus, the development of an early treatment of epithelial and metastatic tumors would offer patients better chances of survival.
Since the YAP transcription coactivator  promotes tumorigenesis and chemotherapy resistance \cite{camargo2007yap1,lamar2012hippo,wang2013mutual,marti2015yap,Warren2018,shen2018hippo}, in principle, controlling its activity could allow a strategic area for therapeutic purposes  in oncology. This could be done in the near future, as the possibility to manipulate defects \textit{in situ} with optical tweezers is now within our reach \cite{bowick2022}. However, the journey to defect-engineered therapies based on YAP deactivation promises to be long, as competing mechanisms \cite{ambivalent} involving defect-induced cell motions have recently been reported to impede malignant cell clearance \cite{zhang2021topological}.

{If currently chemotherapy is very widely used, the efficiency of such long-lasting treatments decreases with time, as cancer cells continually mutate and develop numerous resistance mechanisms \cite{AnnunziataKohn}}. Therefore, there is very often a relapse of patients who initially responded positively to treatment. This relapse occurs around 18 months after the start of chemotherapy in the case of ovarian cancer. A promising alternative, which is less invasive than surgery, is thermal ablation \cite{ChuDupuy}, i.e., killing cancer cells by heat. Different protocols are possible, depending on the heating sources and durations of application. Among them, radiofrequency, microwave, or ultrasound procedure allows for localized heating of cancer cells at temperatures ranging from 50$^\circ$C to 100$^\circ$C for a couple of minutes (at this temperature, the enzymatic activity of the cells stops irreversibly, leading to their death). Another alternative to these methods would be hyperthermia in which the tissue is brought to temperatures neighboring 41 and 45$^\circ$C (via methods such as hyperthermic perfusion or previously aforementioned radiofrequency therapy for instance \cite{SoanesGonder}). Thermal ablation has already proven to be very effective against certain cancerous tumors such as those of the liver. Thermal ablation is also less expensive and requires shorter hospital stays than surgery~\cite{ChuDupuy}. 

Summarizing, thermal ablation of metastic tumors is one among various approaches to treat cancers. As topological defects are correlated to invasive tumor locations, the efficiency of thermal ablation may likely benefit from understanding how heat transfer occurs inside a defective tissue. This calls for the elaboration of a simplified heat diffusion model encompassing the salient features of these complex biophysical phenomena.

\section{Heat propagation is influenced by the presence of topological defects} 

Trefoil and comet-like topological defects of human abdominal  mesothelial (inner membrane) and ovarian  epithelial (outer membrane) cells play a central role in ovarian cancers (about 90\% of ovarian cancers are epithelial tumors).  

However, few in-vivo studies on thermal ablation therapies have been done to date not only for reasons of time and cost but also because of the risks of burning the healthy cells surrounding tumors. Most of the current studies are therefore done by computer modelling and play a major role in the development of treatments \cite{DoostmohammadiAdamerEtAl},
but the field remains largely unexplored and the complexity of the parameters to be taken into account (friction, presence of walls, density fluctuation, turbulence, etc.) does not allow a universal model. 

Here, we elaborate a simple model which allows us to analyse heat propagation in the vicinity of these topological defects.

\subsection{Geometry associated to a disclination of order $m$ in a nematic} 

The geometric theory of defects originates from the pioneering works by Bilby \cite{Bilby1955} and Kr\"oner \cite{Kroner1958}. It is based on the idea that a displacement field $u^i(\textbf{x})$ in elasticity distorts the distance between any pair of points according to:
	\begin{eqnarray}
	\rd l^2&=&\delta_{ij}\rd y^i \rd y^j=\left(\delta_{ij}+\partial_i u_j +\partial_j u_i +\delta_{kl}\partial_i u^k\partial_j u^l\right) \rd x^i \rd x^j,
 \end{eqnarray}
 where the indices $i,j,\ \!\dots$ run from 1 to 3 for the three space coordinates and we assume the standard summation rule on repeated indices.
Hence, everything happens as if the low energy fields propagate on a curved geometry described by the metric 
\begin{equation}
g_{ij}=\delta_{ij}+\partial_i u_j +\partial_j u_i +\delta_{kl}\partial_i u^k\partial_j u^l=\delta_{ij}+2\varepsilon_{ij},
\label{analog-elasticity}
\end{equation}
where $\varepsilon_{ij}$ denotes the strain tensor.

The general form for the metric around a disclination of order $m$ was obtained by S\'atiro and Moraes \cite{SatiroMoraes}. Following an original idea proposed by Joets and Ribotta \cite{JoetsRibotta}, they searched the effective matter-depending geometry leading to the same geodesics as the light paths obtained from the Fermat-Grandjean's principle inside a nematic liquid crystal. Fermat-Grandjean's principle is valid not only for light rays but also for sound rays and a version of it for sound rays in an anisotropic elastic medium (at rest) can be found in \cite{Epstein1992, Babich1994, Ceverny2002}. As nematic liquid crystals have uniaxial optical and acoustic properties along the same principal directions, the optical and elastic Fermat-Grandjean's principles share the same form, and the procedure used by Satiro and Moraes lead to similar effective Riemann geometries, the dielectric constants of the material being replaced by the elastic constants.
The resulting metric in cylindrical coordinates $(\rho,\theta,z)$ is of the following form in terms of the metric tensor $^4g_{\mu\nu}$ [$\mu$~and $\nu$ are space-time indices and run from 0 (time) to 3]:
\be
\rd s^2_{\rm 4D}=^4g_{\mu\nu}\rd x^\mu \rd x^\nu= -c^2\rd t^2+A(\theta)\rd\rho^2+B(\theta)\rho^2 \rd\theta^2+C(\theta)\rho \rd\theta \rd z+\rd z^2, \label{metric}
\ee
with the auxiliary functions
\bea
A(\theta)&=&\cos^2[F(\theta)]+\alpha^2\sin^2[F(\theta)],\\
B(\theta)&=&\alpha^2\cos^2[F(\theta)]+\sin^2[F(\theta)],\\
C(\theta)&=&(1-\alpha^2)\sin[2F(\theta)],
\eea
and $F(\theta)=(m-1)\theta+\varphi_0$. The value of $m$ specifies the topological charge of the defect and for the defects of interest in the context of biological tissues, $m=\pm1/2$. Such geometry describes a static defect: in practice +1/2 defects move at about 1 {\textmu}m/h, therefore the static regime is valid regarding the typical heating times of thermal therapies.  
The case $m=1$ corresponds to ordinary disclination lines which are the liquid crystals analogues of cosmic strings. Such line defects can be formed by either inserting or removing a wedge of material of angle $2\piup(1-\alpha)$ with subsequent identification of the edges. This is the Volterra cut-and-weld process.

\subsection{Heat equation in the presence of topological defects}
The heat equation derives from an enthalpy balance and describes the evolution of the temperature field $T({\bf r},t)$ as \be\partial_tT({\bf r},t)=\lambda\Delta T({\bf r},t),\ee with $\lambda=k/(\mu c)$ the diffusion coefficient given in terms of the thermal conductivity $k$, the density $\mu$ and specific heat $c$ of the material. In the presence of topological defects, one has to modify this equation to account for the non-Euclidean geometry (\ref{metric}).

The ADM formalism (Arnowitt, Deser, Misner, \cite{ADM,Gourgoulhon})
consists in a foliation of spacetime adapted to the Hamiltonian formulation of General Relativity. To do so, the 4D metric $^4g_{\mu\nu}$ ($\mu,\nu=0,1,2,3$) is decomposed into a 3D space metric $g_{ij}$ ($i,j=1,2,3$) through the slicing of the manifold into hypersurfaces orthogonal to the time foliation, $g_{ij}={^4g}_{ij}$, $N=(-{^4g}^{00})^{-1/2}$, $N_i={^4g}_{0i}$ and ${^4g}_{00}=-(N^2-N_iN^i)$. It  implies $\sqrt{-{^4g}}=N\sqrt g$, with $g$ the determinant of the 3-metric and $N$ the so-called lapse function.

This approach was used by Smerlak \cite{Smerlak1,Smerlak29} to write down a heat equation in a foliated spacetime. This requires essentially two modifications: the introduction of the lapse function $N$ in quantities being derived and the use of the Laplace-Beltrami operator
\be\Delta_{\rm LB}(NT)=\frac{1}{\sqrt g}\partial_i[\sqrt g g^{ij}\partial_j(NT)]\ee 
instead of the ordinary Laplacian in a Euclidean manifold. This approach was already used to study heat propagation in the vicinity of topological defects \cite{Manapany22}. In our case, the lapse function $N=1$ and the heat equation becomes, in cylindrical coordinates $(\rho,\theta,z)$
\bea
\frac 1k\frac{\partial T}{\partial t}&=&\Delta_{\rm LB}(T)\nnb\\&=&\Bigl\{\frac{\sin^2[F(\theta)]}{\alpha^2} +\cos^2[F(\theta)]\Bigr\}\frac{\partial^2T}{\partial\rho^2}\nnb\\
&&+\biggl\{\Bigl[
1+(1-\alpha^2)(m-1)
\Bigr]\frac{\sin^2[F(\theta)]}{\alpha^2\rho}
+\Bigl[1-\frac{(1-\alpha^2)(m-1)}{\alpha^2}
\Bigr]\frac{\cos^2[F(\theta)]}{\rho}
\biggr\}\frac{\partial T}{\partial\rho}
\nnb\\&&
-\frac {1-\alpha^2}{\alpha^2\rho}\sin[2F(\theta)]\biggl[\frac{\partial^2T}{\partial\rho\partial\theta}
+(m-1)\frac{\partial T}{\partial\theta}\biggr]\nnb\\
&&+\frac 1{\alpha^2\rho^2}\{\cos^2[F(\theta)]+\alpha^2\sin^2[F(\theta)]\}
\frac{\partial^2T}{\partial\theta^2}
+\frac{\partial^2T}{\partial z^2}
.
\label{eqn:example} 
\eea
The defect geometry couples the radial and angular coordinates $\rho$ and $\theta$, the absence of defect being recovered in the limit $\alpha=1$.

\begin{figure}[!h]
\centering{\includegraphics[height=0.372\columnwidth, angle=0]{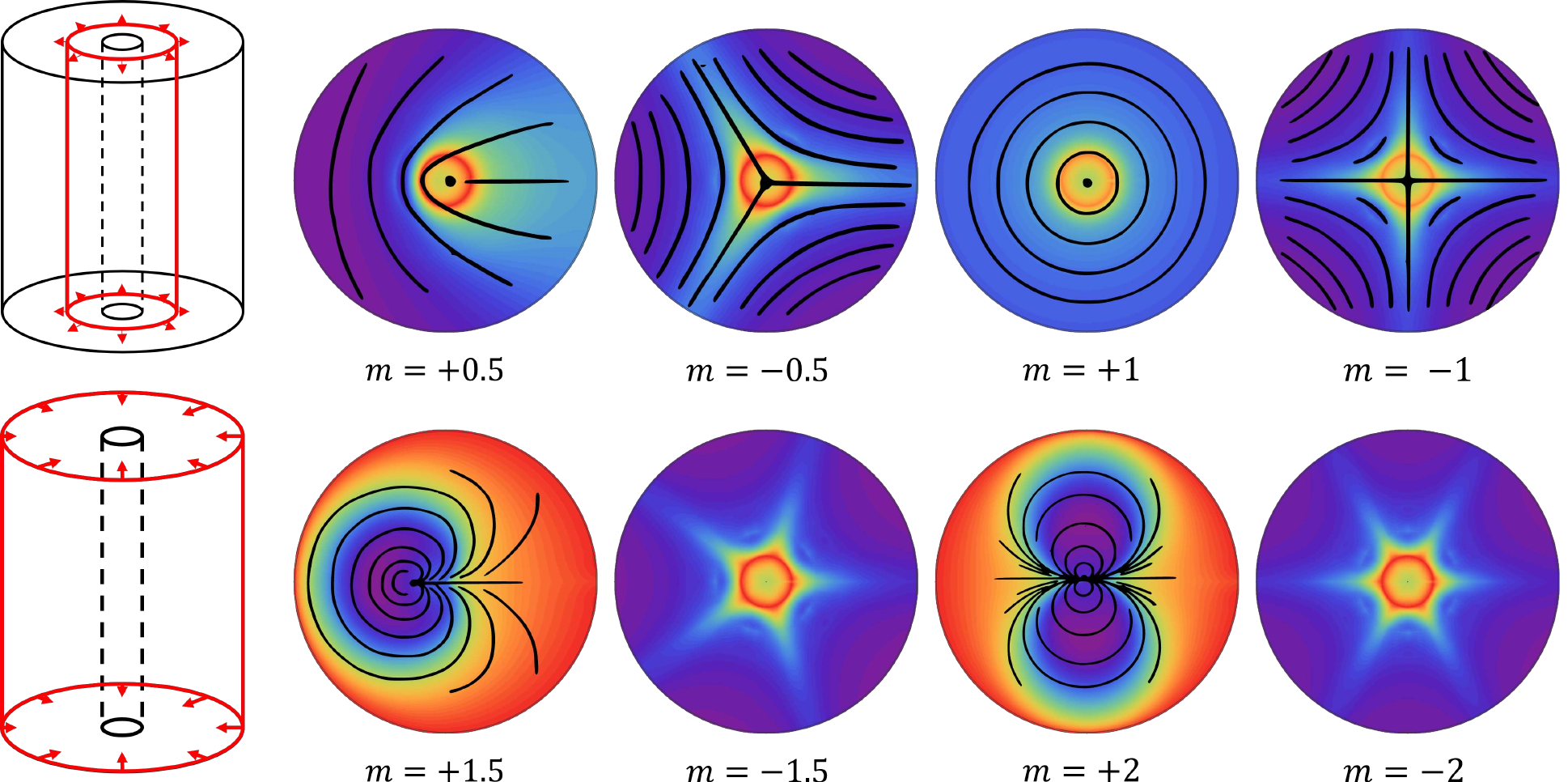}\quad\includegraphics[height=0.38\columnwidth, angle=0]{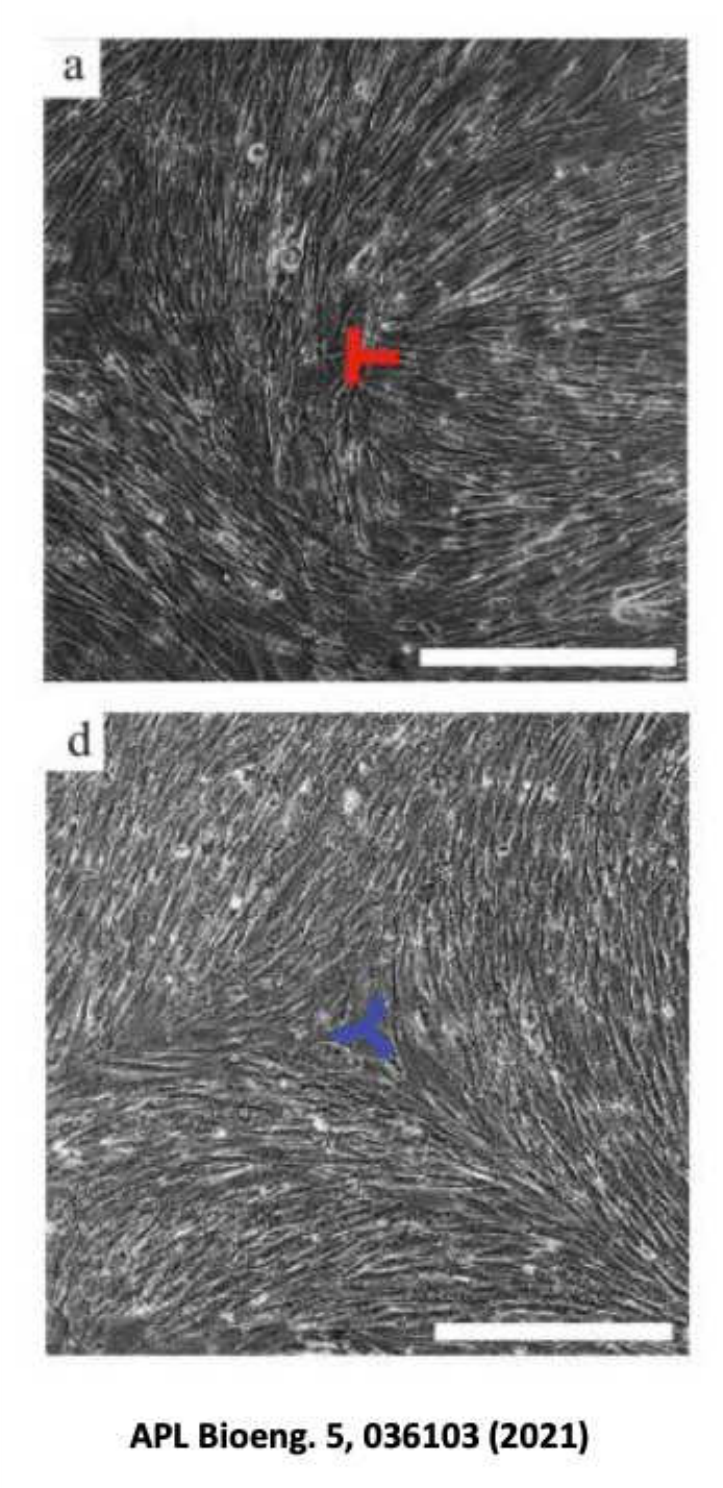}}
\caption[a]{(Colour online) Thermal diffusion (with standard colour code, the red being hotter than the blue) around defects of various topological charges $m$. Here, the disclination parameter is $\alpha=8$. The initial condition sketched on the left shows that heat flux is imposed by fixing the temperature $T_0=1$ on the circular areas shown on the free surface of a finite cylinder (where initially $T=0$) and spreads into space. The different snapshots in the middle show the temperature profile after a finite amount of time. Overlay of the director field (solid lines) with the temperature profile (colors) occurs in any plane perpendicular to the defect axis. The phase contrast images on the right show $m=\pm 1/2$ defects (taken from reference~\cite{zhang2021topological}). 
}
\label{fig1}
\end{figure}

\begin{figure}[!h]
{\includegraphics[width=0.5\columnwidth, angle=0]{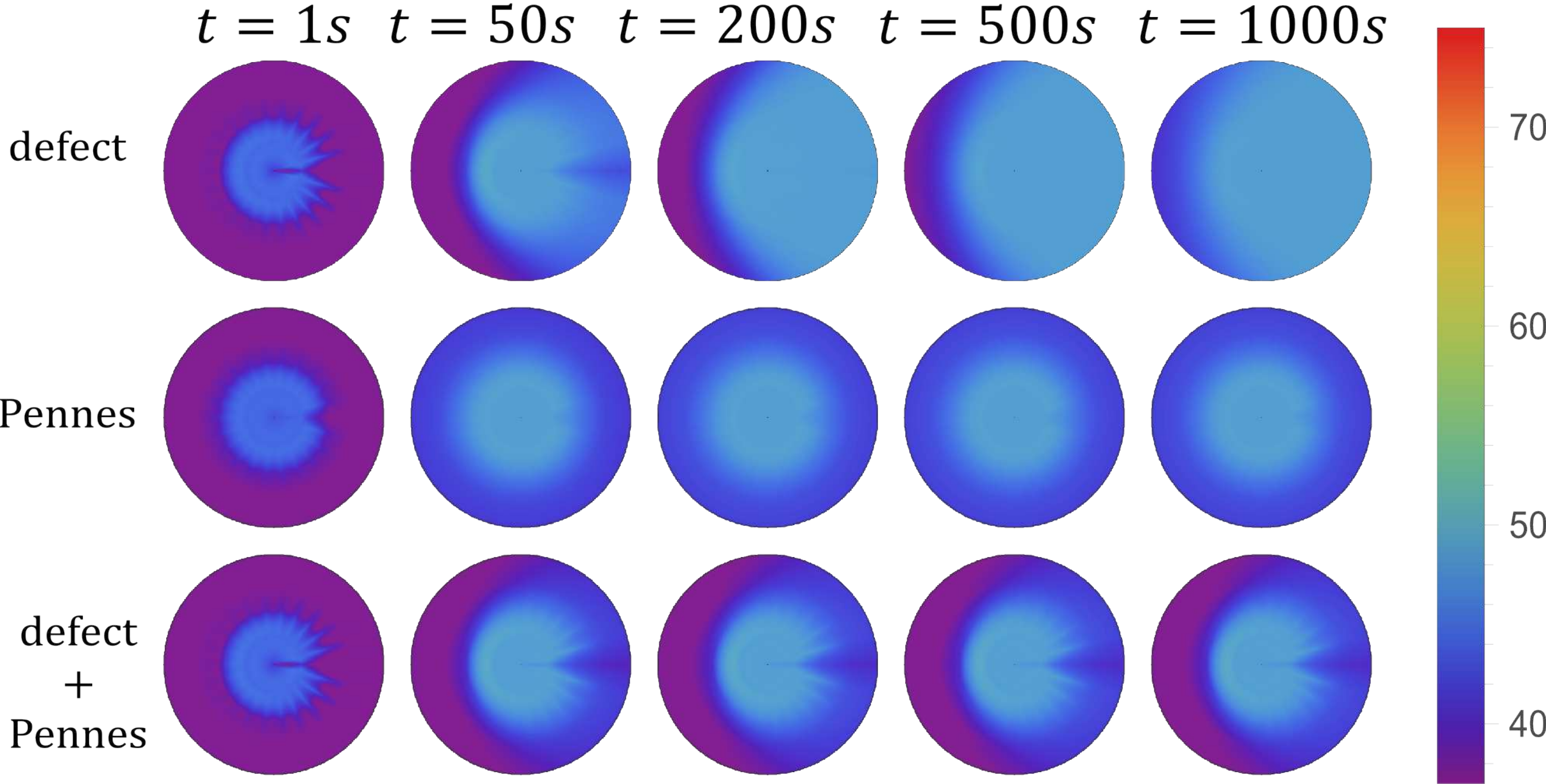}
{\includegraphics[width=0.5\columnwidth, angle=0]{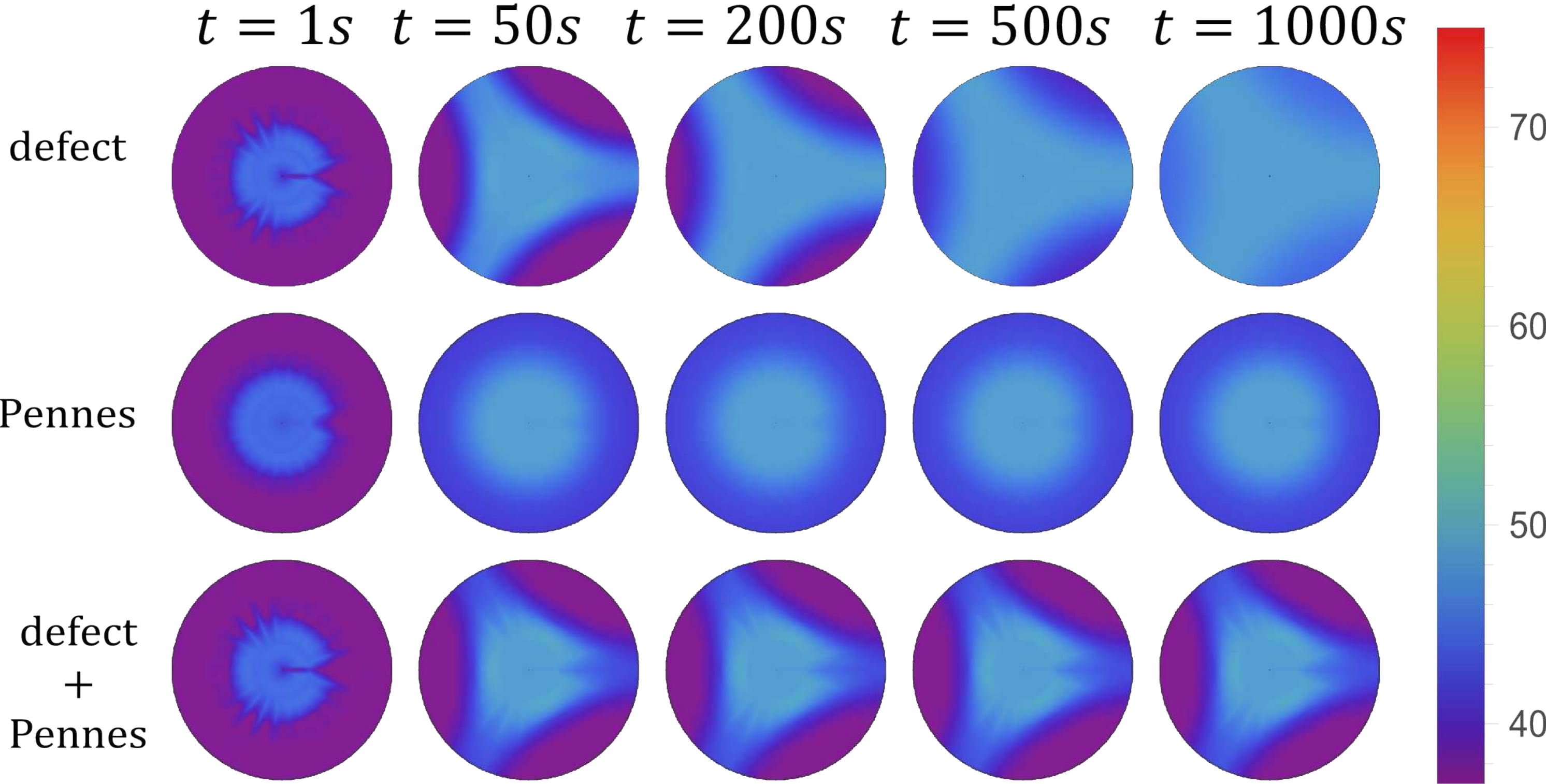}
\includegraphics[width=0.5\columnwidth, angle=0]{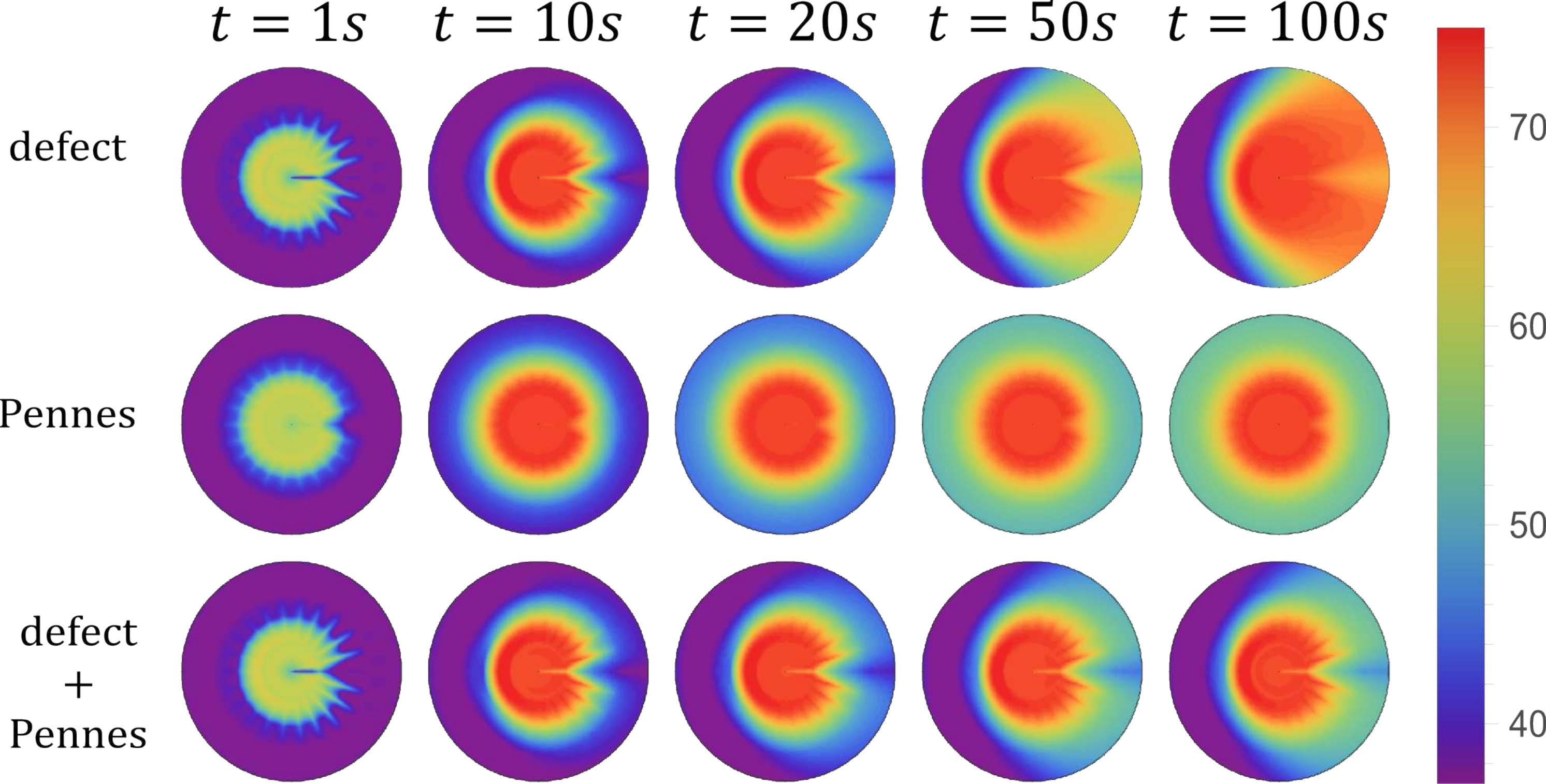}}
\includegraphics[width=0.5\columnwidth, angle=0]{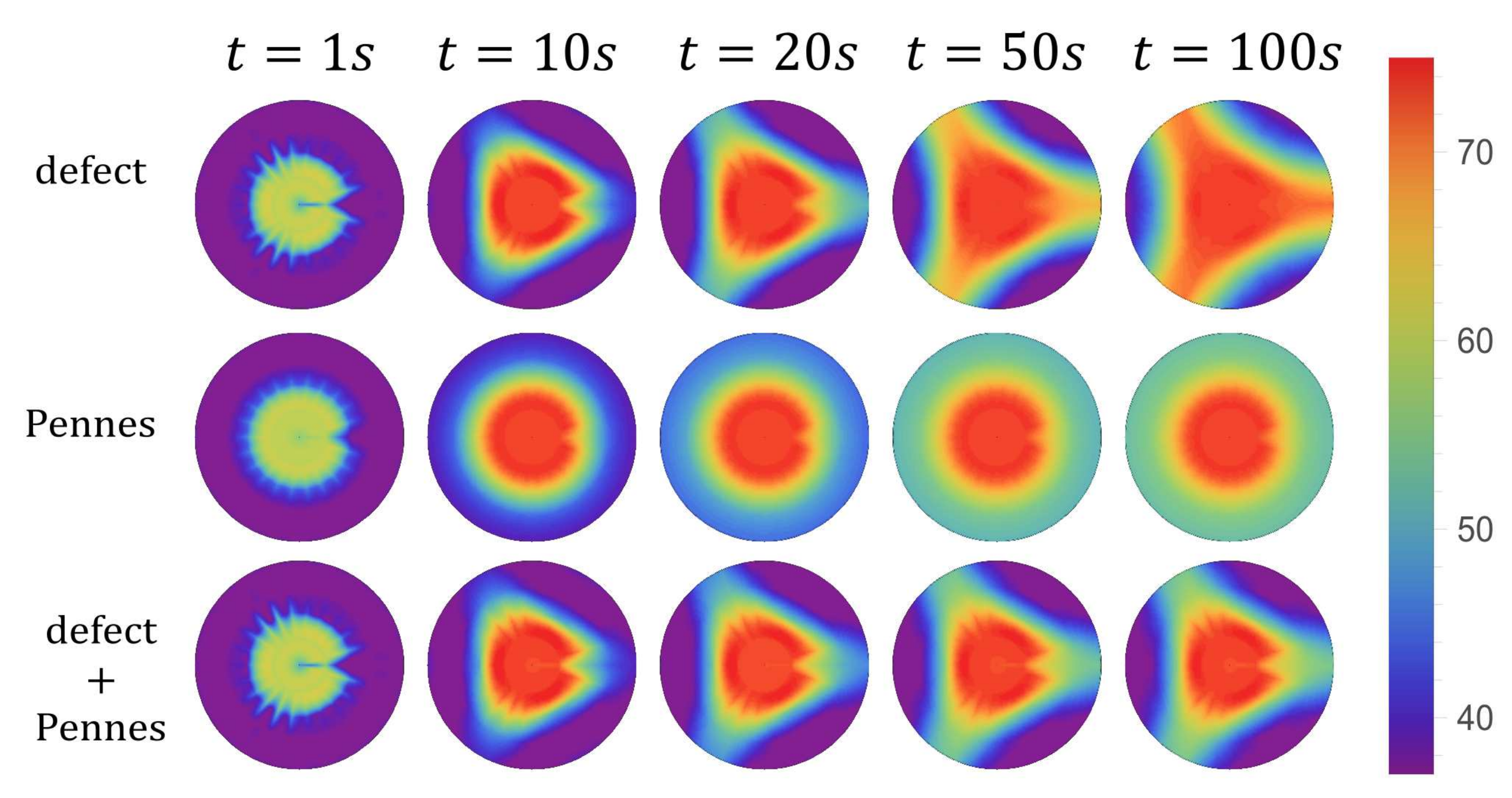}}
\caption[a]{(Colour online) Temperature profiles at various times 
for the comet (left-hand) and trefoil (right-hand) defect  subject to a heat flux at the defect core at a temperature $T_0=46^\circ$C (top) and $T_0=75^\circ$C (bottom). The snapshots are taken at times 
from $t=1$~s to $t=1000$~s from left to right ($T_0=46^\circ$C) and from $t=1$~s to $t=100$~s from left to right ($T_0=75^\circ$C).
The first line shows the temperature profile in the presence only of the topological defect ($\alpha=4$, $m=\pm 1/2$), the second line with the Pennes term only in a purely Euclidean geometry with isotropic temperature profiles, and the third line represents the model of biological tissue with both the topological defect and the Pennes term.
Here, $R_0=5$~mm (burning scale), $R_2=10$~mm (external scale) and the cutoff scale is $R_{\rm cutoff}=0.1$~mm.
}
\label{fig2}
\end{figure}

This equation is then solved numerically using Mathematica. The solutions are derived as interpolation functions using the Finite Element Method. Besides the method, every other parameter such as step size or accuracy digits is chosen automatically depending on the values handled by the solver. The temperature profile in the vicinity of the topological defect is displayed in  figure~\ref{fig1} for various topological charges ranging from $m=-2$ to $m=2$. The material is supposed to be confined within a cylindric domain centered on the defect axis (the core, in the form of a tiny cylinder of radius $R_{\rm cutoff}$, is excluded to avoid numerical divergences) to preserve the axial symmetry.  This  symmetry is compatible with the application proposed in the following sections where  the heat source is produced by an antenna in the shape of a needle. The temperature in this section is fixed at the set value $T_0$ (Dirichlet boundary conditions), either from inside the cylinder or from outside of it (see figure~\ref{fig2}, left).
 The boundary conditions are chosen periodic on the upper and lower walls of the cylinder and in the angular direction.
The temperature profiles corresponding to different times are shown in figure~\ref{fig2}. The cylinder being finite, the asymptotic regime at $t\to\infty$ corresponds to a stationary and uniform temperature profile at $T_0$.
When heating a tissue, a 1/2-defect leaves an unequivocal thermal signature on the temperature profile: in figure~\ref{fig1}, the nematic director field does superimpose to the temperature field in the plane perpendicular to the defect axis. Such results may have important outcomes for diagnosis. Diffusion along the axis is not coupled to the other coordinates and thus is not essential there, so in what follows, we will keep only representations in the perpendicular planes.

\section{Model of heat propagation around topological defects in biological tissues}

Biological tissues are not just inert materials but are rather examples of active matter, i.e., non-equilibrium self-organized systems which do not couple trivially to the energy input. Among them, epithelia are the archetype of active nematic liquid crystals \cite{hirst2017liquid}, each epithelial cell being fuelled by its metabolic activity. The bioheat equation of Pennes \cite{Pennes48} has been introduced to describe arterial temperature in the human forearm. Compared to the ordinary heat equation, it comprises additional terms that include the metabolic contribution (volumic heat source) and the influence of blood flow (convective correction),
\be 
\mu_t c_t\frac{\partial T({\bf r},t)}{\partial t}=k\Delta_{\rm LB} T({\bf r},t)-\mu_bc_b\omega(T-T_a)+q_m. 
\ee 
The l.h.s. corresponds to energy storage (there,  $\mu_t$ is the tissue density [kg/m$^3$] and $c_t$ its specific heat [J/(kg. K)], like $\mu$ and $c$ in the previous section) and the Laplacian  on the r.h.s. (here, $\Delta_{\rm LB}$ indicates the presence of the topological defect) describes as usual the temperature diffusion inside the tissue  (with $k$ the  tissue thermal conductivity [J/s. m. K]).
In biological tissues, there are also at least two sources of energy. The heat transferred by convection by blood to the tissue is described by the second term at the r.h.s. (with $\mu_b$ and  $c_b$ the blood density and specific heat, $\omega$ the local blood perfusion ---  which  ensures the transport of  nutrients and waste products towards and outwards the cells --- and $T_a$  the arterial temperature) and finally, the heat produced by the tissue metabolism given by
the source term, $q_m$.
In the case of a local thermal treatment, an additional term $q_p$ can be introduced as well. Despite its extreme simplicity (the assumption that thermal equilibrium may occur in the capillaries has been questioned) and although several mechanisms are missing from Pennes' equation (the blood velocity field does not appear, the role of thermoregulation, or memory effects are absent, see e.g., \cite{Wissler,AndreozziEtAl,KhaledVafai,ArkinXuHolmes,nabulsi21}), it yet captures, from a limited number of parameters, the salient biophysics of human tissues and is widely used to simulate bioheat transfers at different scales.

 To identify general patterns, we introduce the dimensionless parameters $\bar{r}=\rho/R$ (with $\rho$ the cylindrical coordinate radial distance)  and  $\Theta=(T-T_a)/T_0$ with $T_0$ the setpoint temperature corresponding to a given thermotherapy, and  a typical scale of time 
 \be\tau=\mu_t c_t R^2/k.\ee Neglecting the metabolic term, the Pennes' equation then writes as 
\be \tau\frac{\partial \Theta({\bar{\bold{r}}},t)}{\partial t}=\bar{\Delta}_{\rm LB} \Theta({\bar{\bold{r}}},t)-{\rm Bi}\;\Theta({\bar{\bold{r}}},t), \ee
where the Biot number is 
\be{\rm Bi}=\mu_bc_b\omega R^2/k.\ee

We now compare heat propagation in the inert nematic material to its biological counterpart. We  focus on the cases of comet ($m=+1/2$) and trefoil ($m=-1/2$) defects. Profiles at different times measure the ability of the heat flux to diffuse into the material (figure~\ref{fig2}).  As time increases, warmer areas expand in an anisotropic manner, as a result of the presence of the topological defect (except the isotropic diffusion when $\alpha=1$, i.e., no defect at all in that case).
In figure \ref{fig2}, the first lines show the heat flow in an ordinary nematic in the presence of a topological defect, the second lines correspond to Pennes bioheat equation without any defect and the third lines correspond to the model of interest where the bioheat equation is solved in the presence of the topological defect.

\begin{figure}[!h]
\includegraphics[width=0.51\columnwidth, angle=0]
{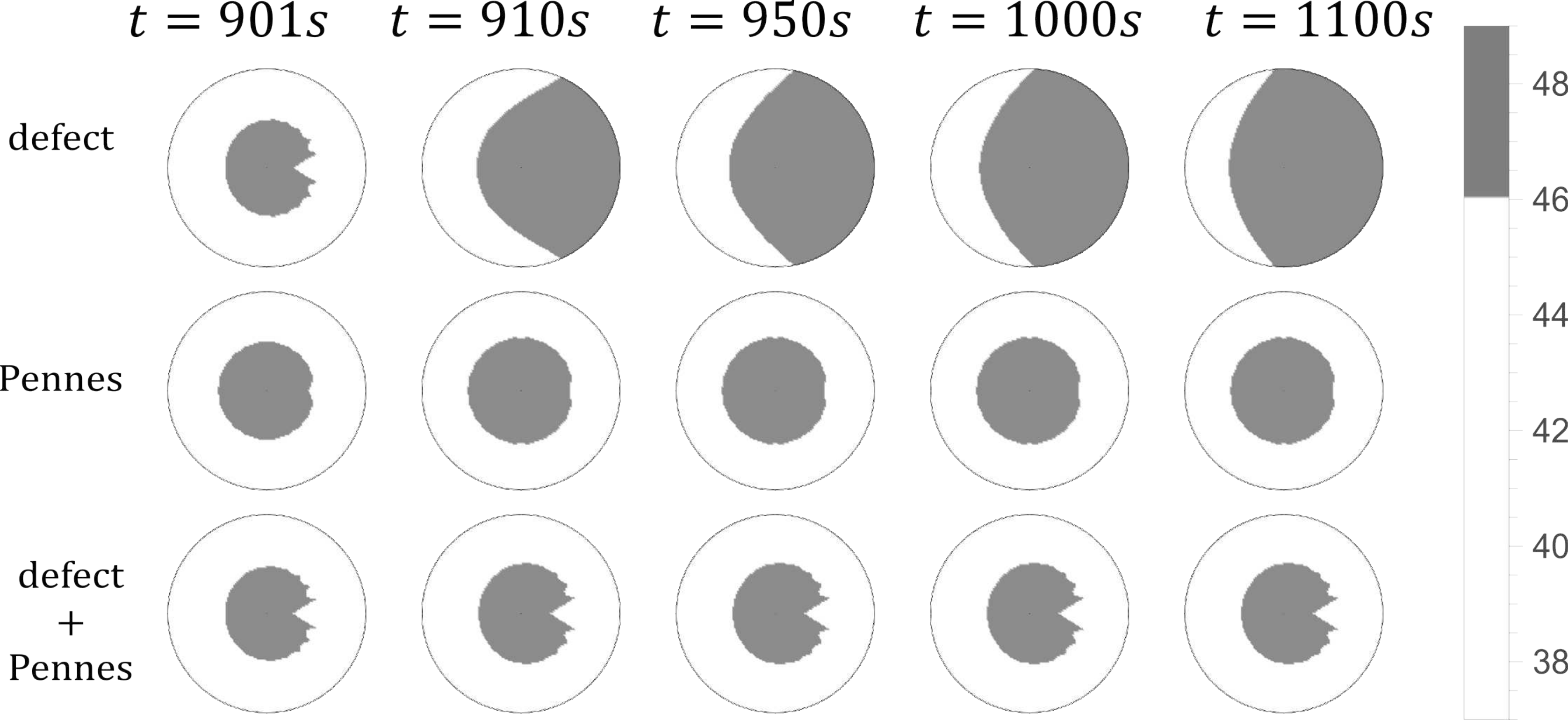}\ \ \includegraphics[width=0.48\columnwidth, angle=0]{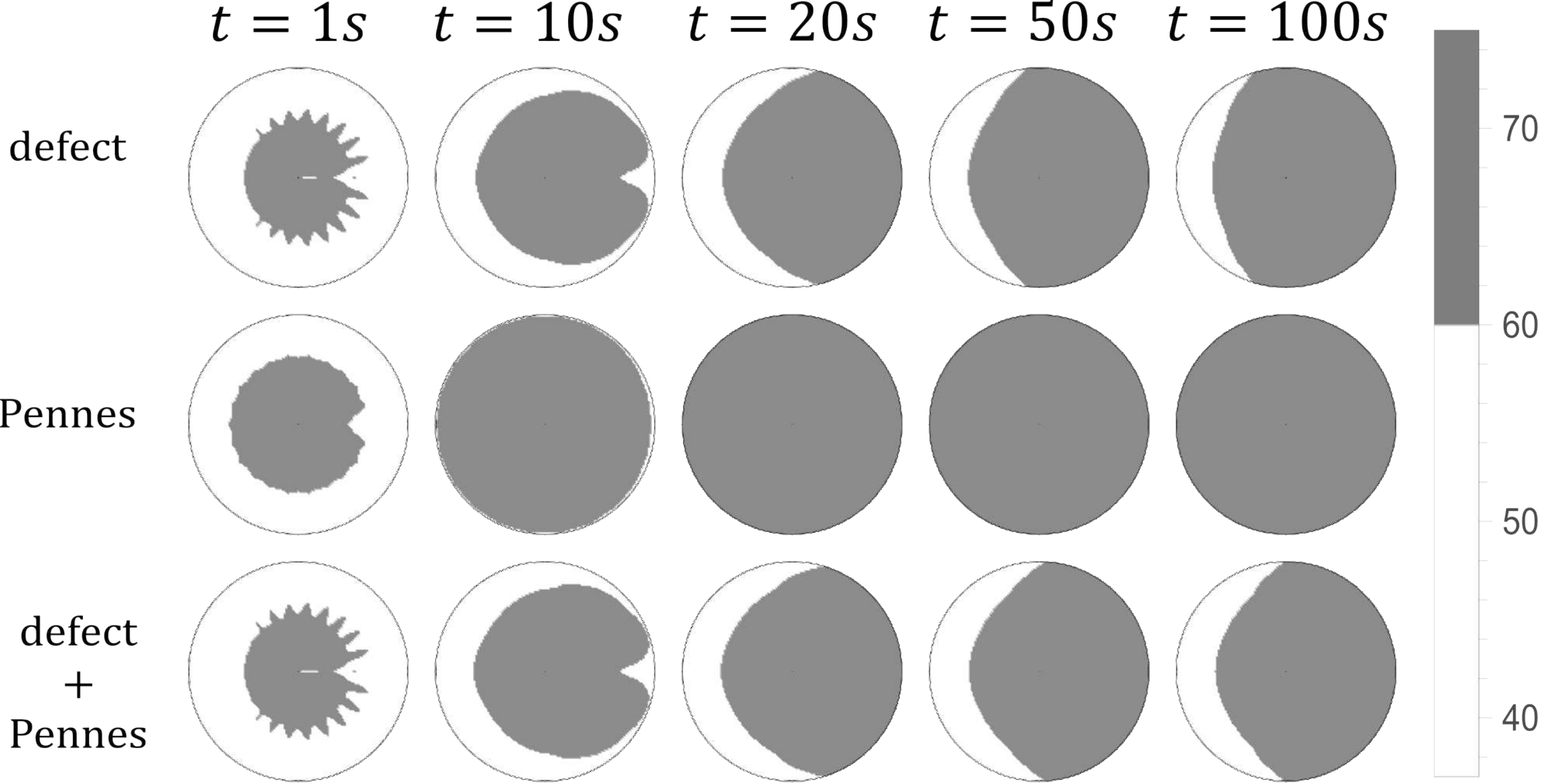}

\caption[a]{Burnt areas for various protocols in the case of a comet-like defect ($\alpha=4$, $m=+1/2$). Left-hand: tissue exposed to a source at $T_0=49^\circ$C. The grey areas at various times represent the part of the system subject to $46^\circ$C during an exposure time of at least $900$~s.
Right-hand: tissue exposed to a source at $T_0=75^\circ$C. The grey areas at various times represent the part of the system subject to $75^\circ$C during an exposure time of at least $1$~s.
The three lines correspond to the same convention as in figures~\ref{fig2}.
}
\label{fig4}
\end{figure}


Several observations are worth noting.  Heat propagates along preferential directions (along three main legs for the trefoil, or one side, say ahead or backward, even though there is no front/rear distinction here).  In all cases, the heat flows in the vicinity of the defect core and 
the temperature in 
the defect core area remains much higher than in the rest of the system.
This is very positive for the application considered here, since ultimately, we will focus on bio-tissues with  defects that concentrate in the core region of  metastatic cells.

In figure~\ref{fig4}  we present in grey the burnt areas, in the same geometrical conditions as in figure~\ref{fig2} for comet defects. Triggering cell apoptosis depends both on the value of the temperature and the time of exposure: typically, cells die when they are exposed at $46^\circ$C during 15 min while they die almost instantaneously when submitted above $60^\circ$C. Thus,
by ``burnt areas'', we mean regions exposed at 
$46^\circ$C during 15 min, or at $60^\circ$C at least during a few seconds, or, by simple interpolation, $55^\circ$C during 400 seconds. Since we found no data at this temperature, that is the simplest approximation we can do. Possible improvements are discussed in the conclusion.

\begin{figure}[!h]
\begin{center}
\includegraphics[width=0.45\columnwidth, angle=0]{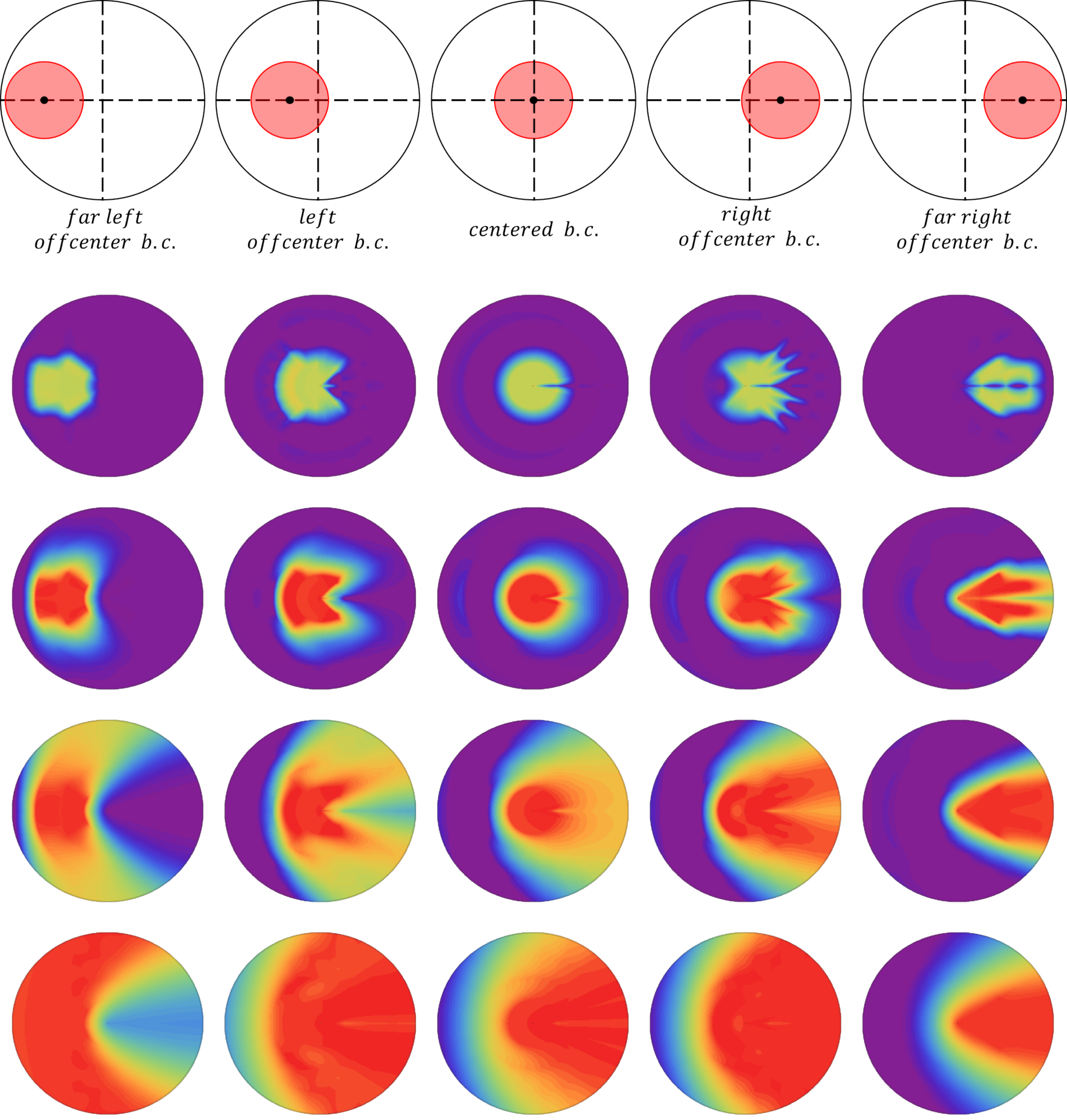}\\ 
\end{center}
\caption[a]{(Colour online) Temperature profiles at various times 
for the comet defect subject to a heat flux off-center from the defect core for the comet defect.
}
\label{fig3bis}
\end{figure}

The next question to address is the efficiency of the different protocols of heating: off-center heating (figure \ref{fig3bis} shows that this is essential to aim at the target precisely), high temperatures for short times (evaporation), lower temperatures applied longer (hyperthermia), and finally a fine balance between temperature level and duration of application (coagulation).
The three situations will be discussed in the next section.

\section{Towards a case-by-case treatment recommendation}

\begin{figure}[!h]
\includegraphics[width=0.31\columnwidth, angle=0]{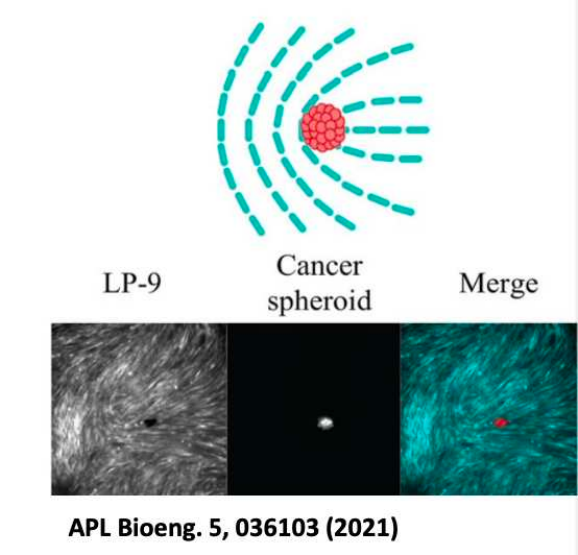}
\includegraphics[width=0.29\columnwidth, angle=0]{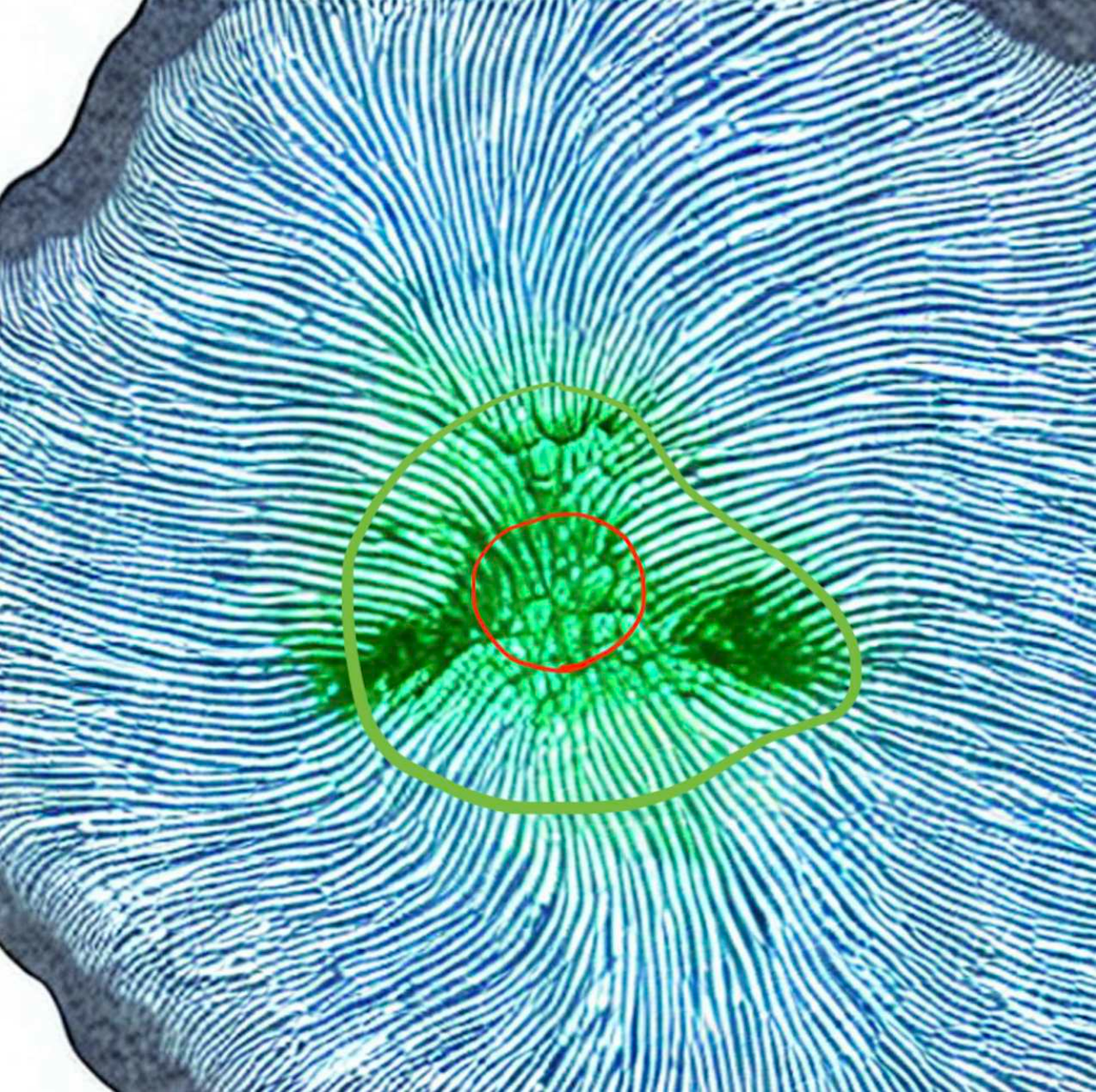}\includegraphics[width=0.38\columnwidth, angle=0]{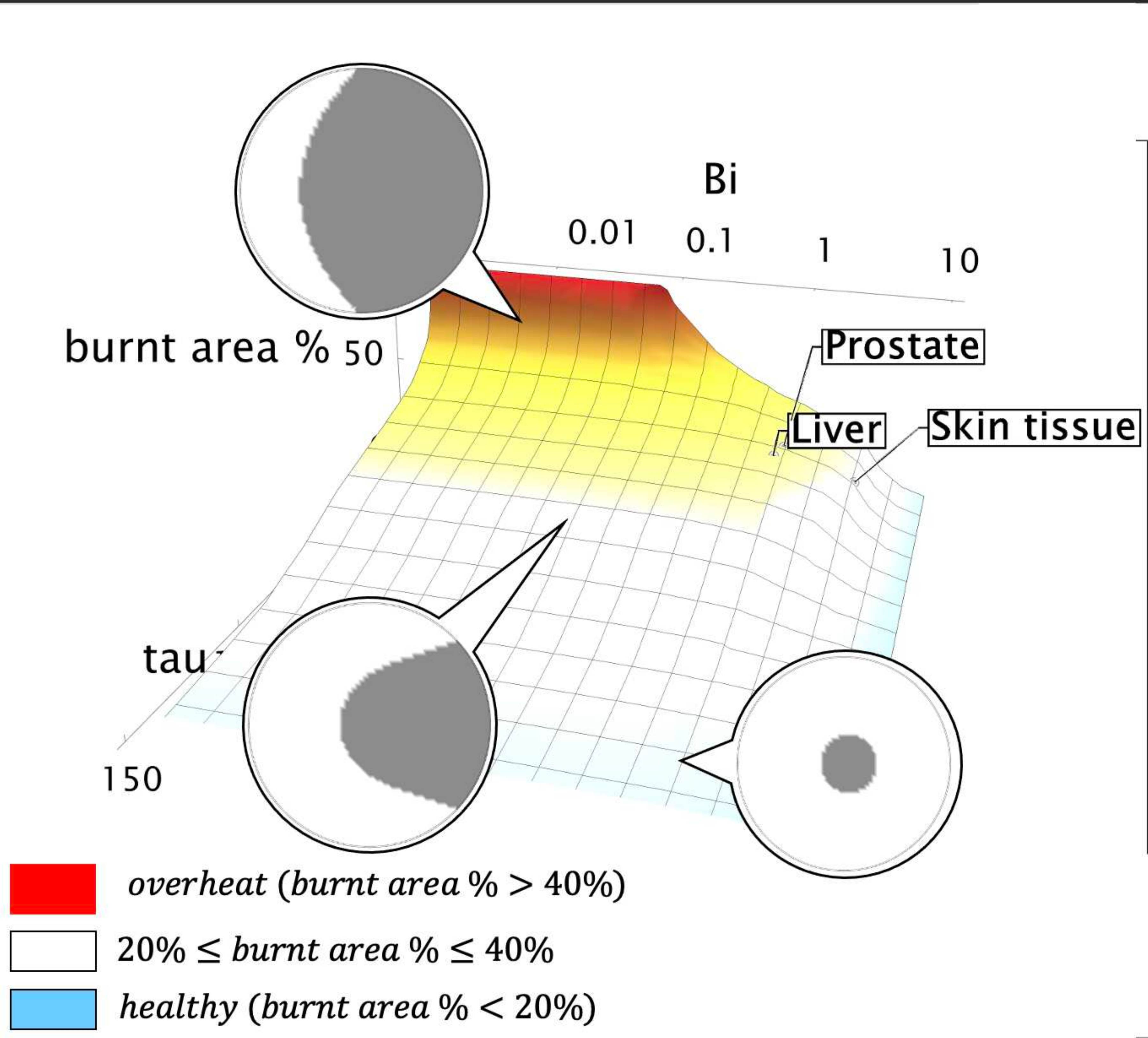}
\caption[a]{(Colour online) Left-hand: An example of tumor located at the center of a comet defect  (taken from reference~\cite{zhang2021topological}).
Center: Sketch (this is an artificial image) of a tumor at the core of a topological defect. The green area represents the tumor of radius $R_t$, and the red contour shows the area, of radius $R_b$, which is submitted to the heating source. For the following geometric parameters $R_b=4$~mm (burning scale), 
$R_2=20$~mm (external scale).
Right-hand:  Evolution of the  relative burnt area in the plane of parameters $\tau$ vs. ${\rm Bi}$. The red part on the surface corresponds to the targeted size burnt for a tumor of size $R_t=10$~mm (tumor scale). Three tissues are located in this parameter space.  }
\label{fig6a}
\end{figure}

We now provide an analysis of the efficiency of thermal ablation protocols for different kinds of biological tissues.
We target by thermal ablation a tumor of radius $R_t$, from a uniform heating protocol on a scale $R_b < R_t$. The tumor is considered as destroyed (burnt) if it has been exposed for a sufficient time, $t_b$, to a temperature at least equal to a value fixed $T_b$, as explained in the previous section. The different protocols allow us to vary the values of $R_b$, $t_b$, and $T_b$. Healthy tissues outside the tumor must be preserved (a sketch is shown in figure \ref{fig6a}).

\begin{table}[h]
	\caption{Parameters of several tissues of the Human body. Note that the blood has tissue-dependent characteristics.}
	\label{tab1}
	\vspace{0.5em}
\begin{center}
\begin{tabular}{cccccc}
	\hline
	Tissue & $\mu_t$  (kg{$\cdot$}m$^{-3}$) & $c_t$ (J{$\cdot$}kg$^{-1}{\cdot}^\circ$C$^{-1}$) & $\mu_b$  (kg{$\cdot$}m$^{-3}$) &  $c_b$ (J{$\cdot$}kg$^{-1}{\cdot}^\circ$C$^{-1}$) & $\omega_b$ (s$^{-1}$) \\
	\hline
	Skin~\cite{YangSun} & 1060 & 2846 & 1000 & 3860 & 0.1  \\
	Liver~\cite{BarnoonAshkiya} & 1079 & 3540 & 1050 & 3639 & $\approx 0.1$   \\
	Prostate~\cite{KalibiTalaee} & 1050 & 3639 & 1060 & 3770 & 0.03 \\
	\hline
\end{tabular}
\end{center}
\end{table}

Each tissue is characterized by the values of the parameters ${\rm Bi}$ and $\tau$ (see table~\ref{tab1} for the biophysical characteristics of different tissues). In the plane of these parameters, we  determine, for a sample of cylindrical tissue of radius $R_2$, the percentage of the burnt area after a certain time has elapsed, for an imposed setpoint temperature fixed at $49^\circ$C, $55^\circ$C and $75^\circ$C, respectively. When this percentage reaches a proportion about 20\% to 40\% of the area of the tissue, a tumor of characteristic radius $R_t\simeq R_2/2$ is considered destroyed ($R_t^2\sim 0.25\ \!R_b^2$). These numbers are chosen for the facility of the representation, but should of course be adapted to the tumor size.

 We have  displayed three different tissues (liver, prostate, and skin) in the ${\rm Bi}-\tau$ plane. These points would represent the target to be reached by the white area.
The position of these targets depends on the biophysical parameters, but also on the size of the tumor via the definition of ${\rm Bi}$ and $\tau$. In the present example, 
$R_t=5$\ \!mm, but for a tumor of half size, the values of $\tau$ and ${\rm Bi}$ would be reduced by a factor of~4.

\begin{figure}[!h]
\begin{center}
\includegraphics[width=0.65\columnwidth, angle=0]{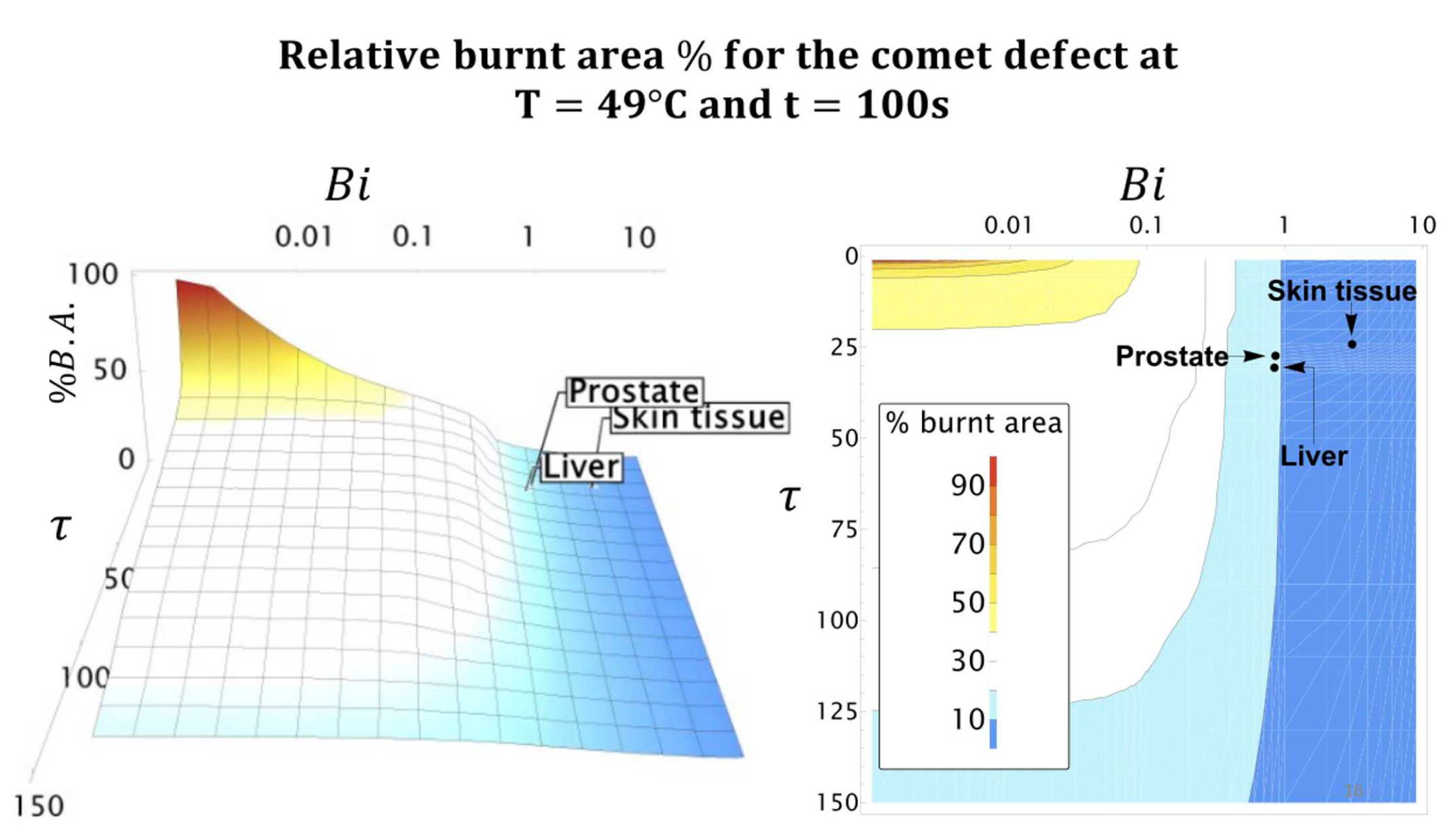}

\includegraphics[width=0.65\columnwidth, angle=0]{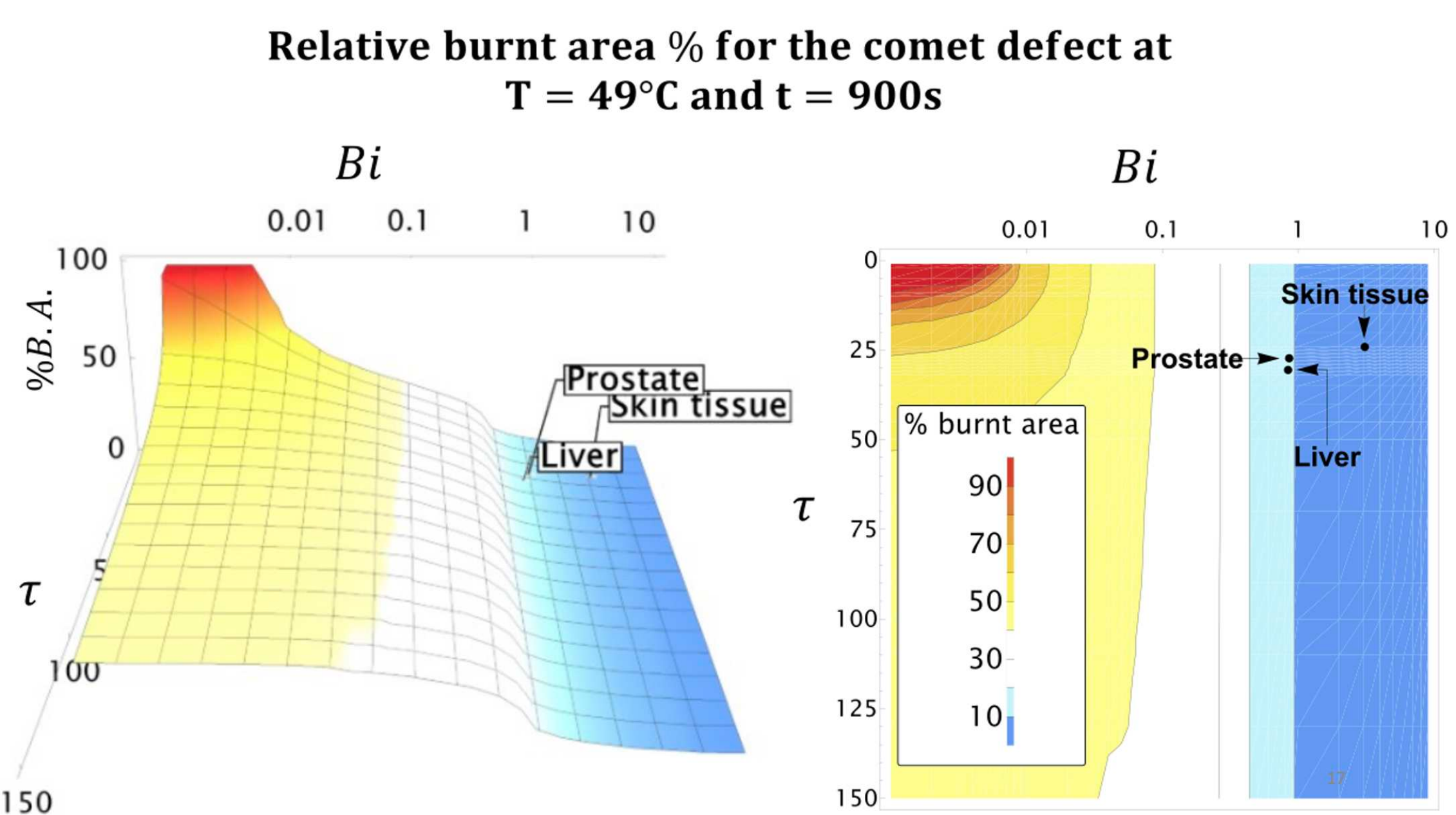}
\end{center}
\caption[a]{(Colour online) Evolution with time of exposure of the  relative burnt area for a tumor of size $R_t=10$ mm heated at 49$^\circ$C. After the 100 s (upper plot), and even after the 900 s, none of the three types of tissues is yet eradicated.}
\label{fig7a}
\end{figure}

Figure~\ref{fig7a} shows our results for comet defects heated at $T_b=49^\circ$C   during $t_b=100$~s and $900$~s. The target is not reached for any of the tissues considered,  since they are not in the white area which corresponds to the expected 20\% to 40\% of the area burnt. In all three cases, the fraction of the area burnt only  covers partially the tumor size.

\begin{figure}[!h]
\begin{center}
\includegraphics[width=0.65\columnwidth, angle=0]{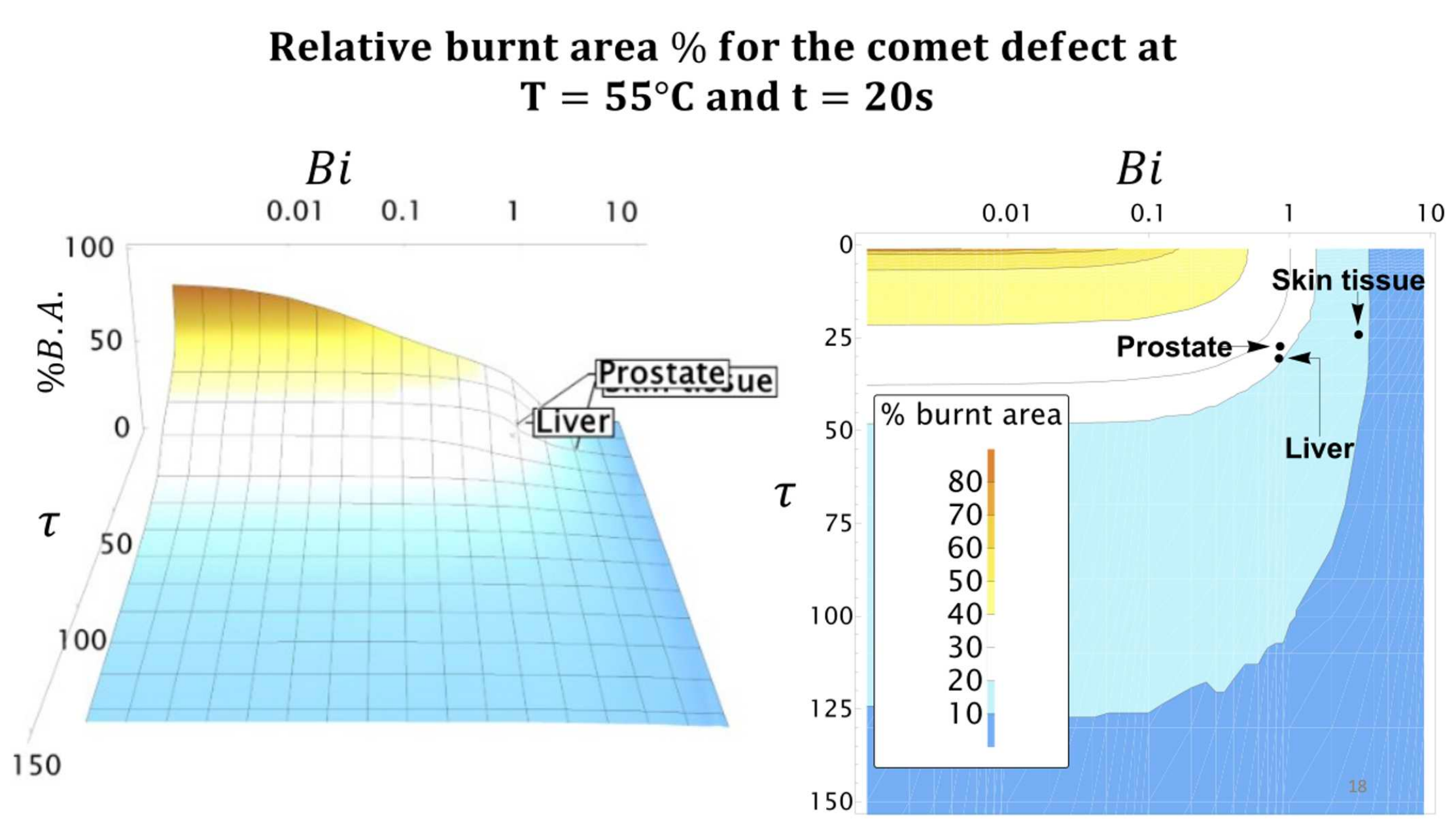}
\end{center}
\caption[a]{(Colour online) Evolution with time of exposure of the  relative burnt area for a tumor of size $R_t=10$ mm heated at 55$^\circ$C during 20 s. Liver and prostate tissues are correctly targeted by these conditions, but the burnt area for skin would not be sufficient to eradicate the tumor.}
\label{fig7b}
\end{figure}

\begin{figure}[!h]
\begin{center}
\includegraphics[width=0.65\columnwidth, angle=0]{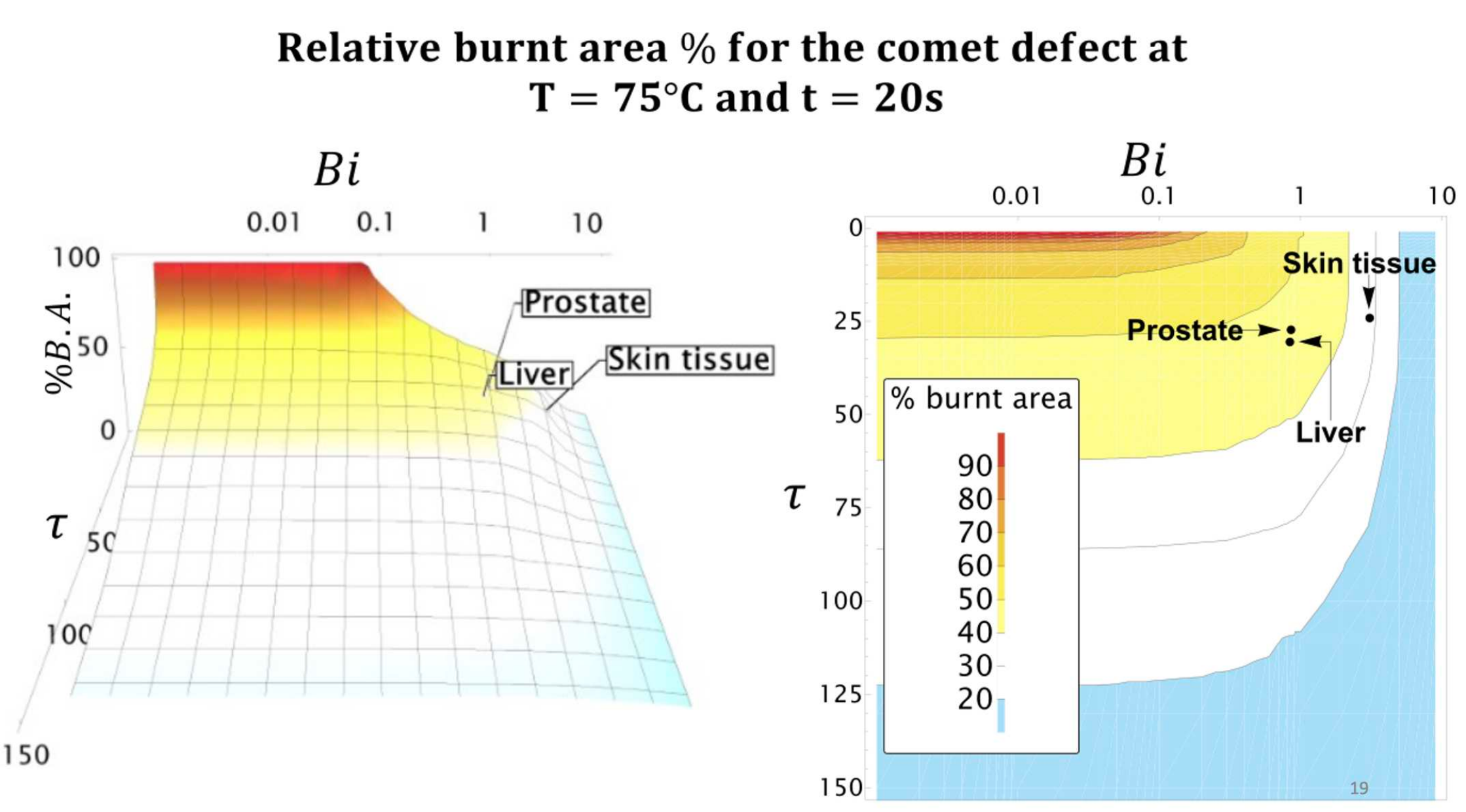}
\end{center}
\caption[a]{(Colour online) Evolution with time of exposure of the  relative burnt area for a tumor of size $R_t=10$ mm heated at 75$^\circ$C during 20 s. Skin tissue is correctly targeted, while a too large fraction of healthy liver and prostate tissues would be burnt by these conditions.}

\label{fig7c}
\end{figure}

At $T_b=55^\circ$C, $t_b=20$~s on the other hand (figure~\ref{fig7b}), the conditions are fulfilled for the liver and prostate tissues, and for the skin, they are satisfied for $T_b=75^\circ$C, $t_b=20$~s (figure~\ref{fig7c}), while at that temperature, the treatment is too aggressive and a significative part of healthy tissue is also destroyed.

Our analysis suggests that prostate and liver tumors should preferably be treated via coagulation protocols (e.g., 55$^\circ$C, 20--50~s), while skin tumors are better treated by evaporation (e.g., 75$^\circ$C, 20~s). Hyperthermia (lower temperatures) in all three cases fails to burn the  tumor to its  whole extent. Although these are very strong statements, we should be cautious with these conclusions which follow from a physical model of a biological tissue, and not from in-vivo studies. The criterion used to decide about the area burnt is also still very approximate, and a careful study of the conditions of apoptosis of cancer cells should be performed to improve the modelization.

\section{Key statements, recommendations, and limitations}
This work introduces a model of heat diffusion in a biological tissue containing half-integer topological defects.  {Comet-like defects ($m=+1/2$) are involved in YAP activity and tumorigenesis, but they may also trap tumor cells and could serve as reliable snitches for image-guided thermal therapies.}
The computations show that the optimum point for tumor burning is very sensitive to the thermophysical properties of the surrounding tissue. They provide a tool to recommend a case-by-case treatment likely to efficiently destroy the tumor cells while preserving the surrounding healthy tissues.

 Several avenues of improvement can therefore be identified to extend the present work.  First, we opted for a cylindrical setting to comply with the defect symmetry, but in practice, this may not reflect the local peculiarities of body organs (yet, changing the setting is not an obstacle in principle and it can be treated from a fully numerical approach).

Second, some of the approximations that we made were due to a lack of knowledge of the properties of the biological tissues (partly due to our ignorance of a part of the literature available, partly because the studies have not yet been performed).  
For example, the dependence of the biophysical properties of in-vivo tissues with temperature and heat damage is not taken into account (recent results evidenced temperature-dependent non-linear changes of the thermal properties of ex-vivo liver~\cite{Guntur,Lopresto1}, brain and pancreas~\cite{MohammadiAsadi} tissues), but there is no theoretical obstacle to incorporate this in the model. 
More delicate would be to account for
heat sink effects of perfusion within in vivo tissue, and those effects are highly non-linear. Living systems dynamically respond to heat sources, dilating blood vessels and increasing interstitial transport to reduce heat build-up.  The thermal damage could also be treated more rigorously using damage models~\cite{Pearce,HaoNourbakhsh}, like Henriques's burn integral~\cite{Pearce} or the Cumulative Equivalent Minutes at 43$^\circ$C dose model (CEM43) which expresses the thermal load on living tissues by estimating the equivalent induced thermal stress in minutes at 43$^\circ$C~\cite{Rhoon}. Finally, improvements to the model should be made in the implementation of the heat source in the tissue. Here, we consider that the heat source is applied {\it locally}, but in a {\it uniform} manner at the inner part of the tissue where the tumor sits, while there should be a penetration law from this inner part to the outer of the tissue. The protocol proposed is also to apply  a constant temperature during time $t_c$, but obvious extensions could be considered, for example cycles of heating-deheating. 

Another improvement of our model consists in taking into account the thermal history of the tumor, as this may affect its tolerance to hyperthermia. Several recent works~\cite{Valentim,Alinei,Vieira}
showed that fractional calculus was a well-suited tool to account for memory effects in biological media.

Despite all the limitations that we have just mentioned and the essential precautions that must be taken with the recommendations to be drawn, this simple model seems very promising to us and can easily be improved to get closer to realistic employment conditions.

\section*{Acknowledgements}
We would like to thank Ralph Kenna (RIP) and Coventry University for supporting Andy Manapany and Leïla Moueddene in this project and their PhDs. Ralph recently fought his own battle with cancer and, while that stymied him from working with us on the scientific part of this project, it did not stop him from securing essential funding that made the project possible. In this way, Ralph's contribution to the fight against cancer endures.  We are all grateful for that of course, but also for much more, for all Ralph has brought to each of us, friendship, advice, and a style of view on life and science.

This work was supported by  the  Coll\`ege Doctoral ``Statistical Physics of Complex Systems'' Leipzig-Lorraine-Lviv-Coventry ($\mathbb{L}^4$). 


\section*{Abbreviations and glossary}

\begin{description}

\item
\emph{Anoikis}:  A form of programmed cell death that occurs in anchorage-dependent cells when they detach from the surrounding extracellular matrix.

\item
\emph{Apoptotic cell extrusion}: Elimination of cells after apoptosis (programmed cell death). See \emph{anoikis}.

\item
\emph{EMT}: Epithelial to mesenchymal transition.

\item
The \emph{epithelium} is a thin protective layer of compactly packed cells that line the outer surfaces of organs and blood vessels throughout the body.

\item
\emph{Extravasion}: Exit of cells in blood or lymphatic vessels. 

\item
\emph{Extrusion}:  Mechanical expulsion of cells from the membrane

\item
\emph{Intravasion}: Entry of cells in blood or lymphatic vessels.

\item
The \emph{mesothelium} is a  membrane that forms the internal lining of several body cavities.

\item
\emph{TACS}: Tumor-associated collagen signatures.

\item
\emph{TAZ}: Transcriptional Co-activator with PDZ-binding Motif (the PDZ domain is a common structural domain of 80-90 amino acids).

\item
\emph{TEAD}: TEA Domain Family Members. The TEA domain is a DNA-binding region of about 66 to 68 amino acids that is named after the two proteins that originally defined the domain: TEF-1 and abaA.

\item
\emph{YAP}:  Yes-Associated Protein,  is a protein that promotes transcription of genes involved in cellular proliferation (this is a component that regulates the organ size, regeneration, and tumorigenesis).

\end{description}



\ukrainianpart

\title{Диференціальна геометрія як можливий шлях для термічної абляції в онкології}
\author{А. Манапані\refaddr{label1,label2},
	Л. Дідьє\refaddr{label1},
	Л. Муддене\refaddr{label1,label2},
	Б. Берш\refaddr{label1,label2},
	С. Фумерон\refaddr{label1} }
\addresses{
	\addr{label1} Лабораторія теоретичної фізики та хімії,  CNRS -- Університет Лотарингії, UMR 7019,
	Нансі, Франція 
	\addr{label2} колаборація $\mathbb{L}^4$, Ляйпциг-Лотарингія-Львів-Ковентрі, Європа
}
\addresses{
	\addr{label1} Лабораторія теоретичної фізики та хімії,  CNRS -- Університет Лотарингії, UMR 7019,
	Нансі, Франція 
	\addr{label2} Колаборація $\mathbb{L}^4$, Ляйпциг-Лотарингія-Львів-Ковентрі, Європа
}
%
%
%

\makeukrtitle

\begin{abstract}
	\tolerance=3000%
	Описується модель терапії гіпертермією, що базується на термодифузії в біологічній тканині, яка має топологічний дефект. Біологічні тканини поводяться як активні рідкі кристали з топологічними дефектами, які, ймовірно, закріплюють пухлини під час метастатичної фази розвитку раку. Завдання терапії полягає в тому, щоб знищити ракові клітини, не пошкоджуючи навколишні здорові тканини. Дефект створює ефективну неевклідову геометрію для низькоенергетичних збуджень, модифікуючи рівняння біотеплопереносу. Проаналізовано та обговорено застосування цього методу для протоколів термічної абляції у випадку різних біологічних тканин (печінки, простати та шкіри).

	\keywords диференціальна геометрія, біотеплоперенос, термічна абляція
	
\end{abstract}

\lastpage
\end{document}